\documentclass[journal,twoside,web]{ieeecolor}
\usepackage{cite}
\usepackage{tmi}
\usepackage{amsmath,amssymb,amsfonts}
\usepackage{algorithmic}
\usepackage{graphicx}
\usepackage{textcomp}
\def\BibTeX{{\rm B\kern-.05em{\sc i\kern-.025em b}\kern-.08em
    T\kern-.1667em\lower.7ex\hbox{E}\kern-.125emX}}
\begin{document}
\title{Implicit Neural Representation for Sparse-view Photoacoustic Computed Tomography}
\author{Bowei Yao, Shilong Cui, Haizhao Dai, Qing Wu, Youshen Xiao, Fei Gao, \IEEEmembership{Member, IEEE}, Jingyi Yu, \IEEEmembership{Fellow, IEEE}, Yuyao Zhang, and Xiran Cai, \IEEEmembership{Member, IEEE}
\thanks{This work was supported by the Shanghai Sailing Program under Grant 21YF1429300. (Corresponding author: Xiran Cai)}
\thanks{Bowei Yao, Shilong Cui, Haizhao Dai, and Qing Wu are with the School of Information Sicence and Technology, ShanghaiTech University, Shanghai
201210, China (e-mail: yaobw@shanghaitech.edu.cn; zengyi2022@shanghaitech.edu.cn; daihz@shanghaitech.edu.cn; wuqing@shanghaitech.edu.cn)}\%thanks{Fis with the School of Information Science and Technology, ShanghaiTech University, Shanghai 201210, China (e-mail: ).}
\thanks{Fei Gao, Jingyi Yu, Yuyao Zhang and Xiran Cai are with the School of Information Science and Technology and the Shanghai Engineering Research Center of Intelligent Vision and Imaging, ShanghaiTech University, Shanghai 201210, China (e-mail: gaofei@shanghaitech.edu.cn, yujingyi@shanghaitech.edu.cn, zhangyy8@shanghaitech.edu.cn, caixr@shanghaitech.edu.cn).}
}

\maketitle

\begin{abstract}
High-quality imaging in photoacoustic computed tomography (PACT) usually requires a  high-channel count system for dense spatial sampling around the object to avoid aliasing-related artefacts.
To reduce system complexity, various image reconstruction approaches, such as model-based (MB) and deep learning based methods, have been explored to mitigate the artefacts associated with sparse-view acquisition.
However, the explored methods formulated the reconstruction problem in a discrete framework, making it prone to measurement errors, discretization errors, and the extend of the ill-poseness of the problem scales with the discretization resolution.
In this work, an implicit neural representation (INR) framework is proposed for image reconstruction in PACT with ring transducer arrays to address these issues.
Specially, the initial heat distribution is represented as a continuous function of spatial coordinates using a multi-layer perceptron (MLP).
The weights of the MLP are then determined by a training process in a self-supervised manner, by minimizing the errors between the measured and model predicted PA signals.
After training, PA images can be mapped by feeding the coordinates to the network.
Simulation and phantom experiments showed that the INR method performed best in preserving image fidelity and in artefacts suppression for the same acquisition condition, compared to universal back-projection and MB methods.
In the experimental data, the INR method improved signal-to-noise-ratio (contrast-to-noise-ratio) by 6.83–19.42 dB (5.85–19.33 dB), compared to the other methods.
These results clearly demonstrated the value of INR for high-quality PACT image reconstruction with sparse data and its potential in reducing the complexity of PACT systems.
\end{abstract}


\section{Introduction}
\label{sec:introduction}
\IEEEPARstart{P}{hotoacoustic} computed tomography (PACT) combines the high optical contrast with the high penetration depth of acoustic imaging, to enable high-resolution functional imaging of tissues in the body\cite{xu2006photoacoustic, wang2009multiscale}. 
PACT has been widely applied in preclinical studies for whole-body imaging of small animals, to map the vascular networks and the whole-body hemodynamics~\cite{b4,b5,yeh2017dry, choi2023deep}, to study placental functions~\cite{zhu2024longitudinal} and to explore resting-state functional correlation in the mice brain~\cite{b3, zhang2018high}.
In human imaging, PACT has been applied for functional brain imaging\cite{na2022massively}, the diagnosis of vascular diseases\cite{b6,b7} and cancer detection\cite{lin2018single}.
In PACT, the objective is to recover the initial pressure distribution initiated by short-pulsed (nanosecond duration) laser illumination on the target\cite{wang2016practical}. After absorbing the energy of pulsed light, the absorbers in the tissue undergo thermal expansion, which then generate sound pressure, i.e. photoacoustic (PA) signal, propagating outward.
Then, the PA signals are received by ultrasonic transducers positioned at different locations outside the tissue, and are used to reconstruct the image representing the distribution of the absorbed light’s energy in the tissue using algorithms, such as time reversal (TR)\cite{treeby2010photoacoustic}, universal back-projection (UBP)\cite{b12}, or model-based iterative methods\cite{b10,Huang2013Full-Wave}.

High quality PACT imaging requires a dense spatial sampling around the object to satisfy the Nyquist sampling theorem and to avoid spatial aliasing related artefacts\cite{Hu2020Spatio}.
Thus, a high-channel count acquisition system and an array of large number of transducer elements are usually deployed for imaging, resulting in increased cost and complexity of the PACT system.
In sparse-view PACT, to eliminate the artifacts in the images induced by sparse data, such as the streak-type artifact, and improve image quality, various image
 reconstruction approaches have been explored.
Sacrificing spatial resolution of the image, the PA signals may be low-pass filtered so that the required spatial sampling frequency may be lowered, as the wavelength used by the UBP method for image reconstruction is increased\cite{Hu2020Spatio,b5}.
Image quality in sparse-view PACT can be improved with MB methods which formulate image reconstruction as an inverse problem.
This inverse problem is often solved by optimization, i.e., minimizing the error between the measured PA signals and the theoretical ones predicted by a certain PA forward model\cite{b10,Huang2013Full-Wave}.
However, the inverse problem is a highly ill-posed and unstable problem which is prone to the errors in measurements and modeling.
Therefore, applying proper regularization to constrain the problem to relieve the ill-poseness is often necessary\cite{dong2015algorithm}. 
Nevertheless, under sparse sampling conditions, the ill-poseness of the inverse problem deteriorates.
Thus, the inverse problem usually requires more restricted regularization which heavily affects the image quality and the inversion may be even impossible\cite{gutta2018accelerated}.
Deep learning based methods have been also introduced to improve image reconstruction quality from sparse data, using the networks based on U-net structure\cite{tong2020domain ,guan2020limited, davoudi2019deep} or attention-steered network\cite{guo2022net} trained in a supervised manner.
The training process requires a large dataset which is generally difficult to obtain in PACT and the generalization capability of the method heavily depends on the quality of the dataset.
Integrating a diffusion model to the inversion in the MB method, the optimization problem for image reconstruction with sparse data can be better constrained\cite{song2023sparse}.
However, the diffusion model essentially learns the prior information of the data distribution to better constrain data consistency.
It may only be applicable to the learnt data distribution while not to other scenarios.
In addition, the training process to learn the data distribution with diffusion models takes hours\cite{song2023sparse}.
Thus, for distinctly different data distribution conditions, it cannot reconstruct the image in a timely manner.

A common feature in all the currently adopted strategies to overcome the ill-pose problem of the inverse problem in PACT, particularly under sparse-view, is that the inversion is formulated in a discrete framework, i.e., the images to be optimized and the forward model are represented as discrete matrix.
Thus, the reconstruction is prone to measurement errors and discretization errors at low resolution.
At high-resolution, the increased number of unknowns imposes high time and space complexity and makes the inverse problem more ill-posed.

In recent years, a new paradigm to formulate the inverse problem has emerged in the computer vision and graphics communities\cite{b17}. 
Using implicit neural representation (INR) models, the interested unknowns are represented as continuous functions, whose parameters are obtain after network training in a self-supervised manner.
Thanks to the continuity imposed by INR, this approach has achieved superior performance for various computer vision tasks\cite{b17,zhang2020nerf++,muller2022instant}, and has been translated to medical image reconstruction problems\cite{lihonggen2021,wuqing2023,fengruimin2023}.
Exemplary applications include recovering artifact-free images from sparse-view sinogram data in CT \cite{shen2022nerp,sun2021coil,wu2023self}, as well as improving image reconstruction performance under both confocal and non-confocal settings in non-line-of-sight imaging\cite{shen2021non}, etc.


Enlightened by the aformentioned studies, in this work, we propose a framework based on INR for PACT image reconstruction with sparse data acquired by ring transducer arrays.
Specifically, the initial heat distribution is represented implicitly with a neural network whose parameters are determined by the radio frequency (RF) PA signals after training.
The training process essentially involves minimizing the errors between the measured PA signals and the ones predicted by a forward model relating the neural representation of the initial heat distribution to PA signals in a self-supervised manner.
After the training, the PA images can then be mapped by feeding the coordinates to the network.
In the following, we present, evaluate and validate the proposed framework by comparing it with UBP and MB methods, using both simulated and experimental data of different sparsity.
The results show that the proposed method performed best in terms of image fidelity and artifacts suppression than the reference methods for the same sparse-view sampling conditions.

\section{Materials and Methods}
\subsection{Photoacoustic Wave Equation}
In PACT, with nanosecond laser pulses, the thermal expansion of the irradiated region is not affected by its neighboring region\cite{xu2006photoacoustic}.
The temporal profile of the light source is appreciated as a Dirac delta function and the pressure of the generated ultrasonic waves in a homogeneous medium is given by~\cite{dean2012acceleration}
\begin{equation}
\frac{\partial^{2} p(\boldsymbol{r},t)}{\partial t^{2}} - c^{2} \nabla ^{2} p(\boldsymbol{r},t) = \Gamma H(\boldsymbol{r}) \frac{\partial \delta(t)}{\partial t}\label{eq1}
\end{equation}
where $c$ is the speed-of-sound (SoS) of the medium, $\Gamma$ is the Grueneisen parameter and $H(\boldsymbol{r})=\mu_{a}(\boldsymbol{r}) U(\boldsymbol{r})$ is the amount of energy absorbed per unit volume at position $\boldsymbol{r}=(x,y)$ with $\mu_{a}(\boldsymbol{r})$ the optical absorption coefficient and $U(\boldsymbol{r})$ the light fluence, and $p(\boldsymbol{r},t)$ represents the pressure at $\boldsymbol{r}$ and time $t$. Considering the initial conditions
\begin{equation}
    p(\boldsymbol{r},t)|_{t=0} =  \Gamma H(\boldsymbol{r})\label{eq2}
\end{equation}
\begin{equation}
   \frac{\partial p(\boldsymbol{r},t)}{\partial t} \bigg| _{t=0} = 0 \label{eq3}
\end{equation}
 Eq. (\ref{eq1}) can be solved by Green's function method and $p(\boldsymbol{r},t)$ can be expressed as\cite{dean2012acceleration}:
\begin{equation}
    p(\boldsymbol{r},t) = \frac{1}{4\pi c^{2}}\frac{\partial }{\partial t}\int_{S(t)} \frac{p(\boldsymbol{r'})}{|\boldsymbol{r}-\boldsymbol{r'}} d S(t)\label{eq4}
\end{equation}
where $S(t)$ represents the spherical wavefront originated from $r'$ with a radius of $ct$. 
\begin{figure*}
\centering
\includegraphics[width=\textwidth]{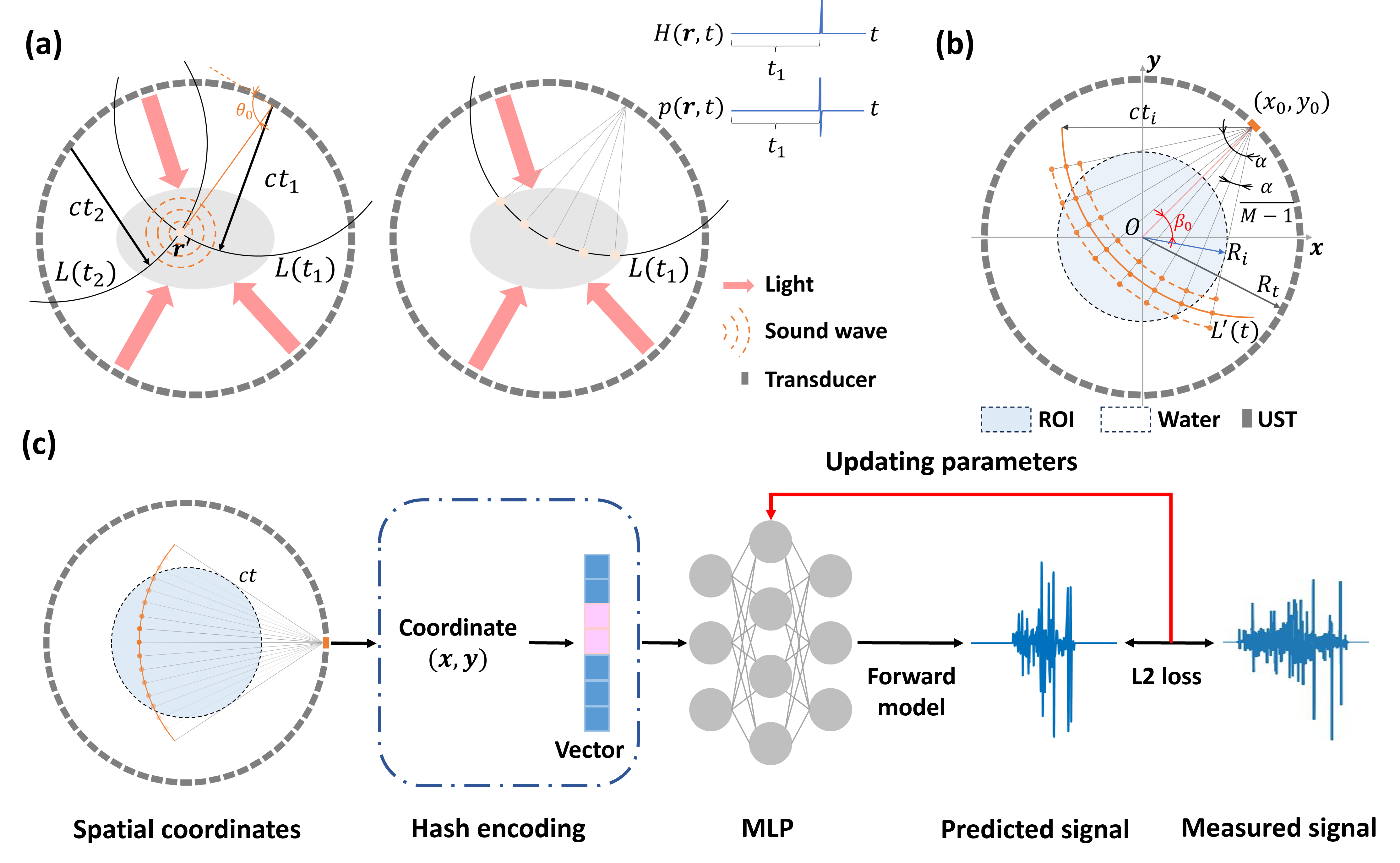}
\caption{Overview of the neural representation reconstruction for PACT. (a) Schematic demonstration of the propagation of the sound wave excited by the laser. (b) Spatial-temporal discretization for the PA forward model. Points on the orange arcs represent the sampled coordinates for calculating the PA signals with the forward model. (c) The flowchart of the proposed INR method. ROI, region of interst; UST, ultrasound transducer.}
\label{Fig1}
\end{figure*}

\subsection{Universal Back-projection Method}
In 2D PACT with a ring transducer array of radius $R_{t}$ (Fig.~\ref{Fig1}), image reconstruction can be treated as inverting Eq.~(\ref{eq4}), i.e. reconstruct the initial pressure $p(\boldsymbol{r'})$ from the RF data $p(\boldsymbol{r},t)$ collected by the ring array.
The UBP method to solve the inverse problem is expressed as~\cite{b12}:
\begin{equation}
    p(\boldsymbol{r'})= \frac{1}{\Omega_{0}}\boldsymbol{\int}_{S_{0}}\big[2p(\boldsymbol{r},t) - \frac{2t\partial p(\boldsymbol{r},t)}{\partial t}\big]\frac{\cos{\theta_{0}}}{|\boldsymbol{r}-\boldsymbol{r'}|^{2}}dS_{0}
    \label{eq5}
\end{equation}
where $\theta_{0}$ is the angle between the vector pointing to the source position $r'$ and transducer surface $S_{0}$ (Fig.~\ref{Fig1}(a)).
${\Omega_{0}}$ is a solid angle of the $S_{0}$ with respect to the wave source.
For transducer elements of planar geometry, ${\Omega_{0}} = 2\pi$, and ${\Omega_{0}} = 4\pi$ for transducer elements of cylindrical geometry used in this work. 
Under the condition of full and dense view sampling, accumulating the projection data simply with Eq.~(\ref{eq5}) yields images with good quality, otherwise, artifacts persist in the reconstructed images.

\subsection{Model-based Method}
MB methods have been also applied in PACT for image reconstruction, by solving an inverse problem iteratively with a defined forward model of PA signals~\cite{b10,Huang2013Full-Wave}.
For numerical evaluation of the forward model, omitting the constant factors, the derivative in Eq.~(\ref{eq4}) is approximated as~\cite{dean2012acceleration}: 
\begin{equation}
    p(\boldsymbol{r},t) = \frac{I(t+\Delta t) - I(t-\Delta t)}{2\Delta t}\label{eq6} 
\end{equation}
where $\Delta t$ is a time interval for derivative calculation which is much smaller than the sampling interval of of PA signal, with $I(t)$ defined as
\begin{equation}
    I(t) =  \int _{L'(t)} \frac{H(\boldsymbol{r'})}{|\boldsymbol{r-r'}|}dL'(t)\label{eq7}
\end{equation}
$I(t)$ is discretized as
\begin{equation}
    I(t) \approx  \frac{1}{2}\sum _{l=1} ^{M-1} \Big [\frac{H(\boldsymbol{r_{l}'})}{|\boldsymbol{r-r_{l}'}|}+\frac{H(\boldsymbol{r_{l+1}'})}{|\boldsymbol{r-r_{l+1}'}|}\Big]d_{l,l+1}\label{eq8}
\end{equation}
where $L'(t)$ is uniformly sampled by $M$ points across the angular sector covering the circular region-of-interest (ROI) of radius $R_{i}$ (Fig.~\ref{Fig1}(b)).
Assuming uniform SoS distribution of the medium, $L'(t)$ is an arc of a circle centered on the position of the transducer element.
Thus, the segment $d_{l,l+1}$ of curve $L'(t)$ can be expressed as: 
\begin{equation}
    d_{l,l+1} = \left\{ \begin{aligned}
        &\frac{\alpha}{M-1}ct & 1 \leq l \leq M-1 \\
        &  0 & l=0, l=M\\
    \end{aligned}
    \right. \label{eq9}
\end{equation}
where $l$ indexes the segments, and $\alpha$ represents the opening angle of the sector covering the whole ROI viewing from the transducer element, calculated as:
\begin{equation}
    \alpha = 2 arcsin(\frac{R_{i}}{R_{t}})\label{eq10}
\end{equation}
Eq.~(\ref{eq8}) can be reduced to~\cite{dean2012acceleration}
\begin{equation}
    I(t) \approx  \frac{1}{2}\sum _{l=1} ^{M} \Big [\frac{H(\boldsymbol{r_{l}'})}{|\boldsymbol{r-r_{l}'}|}\Big](d_{l-1,l}+d_{l,l+1})\label{eq11}
\end{equation}
Combining Eq.~(\ref{eq6}) and Eq.~(\ref{eq11}), the forward model (Eq.~\ref{eq4}) can be formulated in matrix form as:
\begin{equation}
    \textbf{p} = \textbf{A}\textbf{H}\label{eq12}
\end{equation}
where $\textbf{p}$ is the model predicted PA signals collected by the transducers, $\textbf{A}$ is the measurement matrix, and $\textbf{H}$ is the vectorized image to reconstruct.
With the MB method, PA images $\textbf{H}$ are then reconstructed by minimizing the loss function $\mathcal{L_{\textbf{H}}}$ between $\textbf{p}$ and experimentally measured PA signal $\textbf{p}_{m}$ by the ring array, 
\begin{equation}
   \mathcal{L_{\textbf{H}}} = ||\textbf{p}_{m}-\textbf{A}\textbf{H}||_{2}^{2} + \lambda R(\textbf{H})\label{eq13}
\end{equation}
where $R(\textbf{H})$ and $\lambda$ are the regularization term and regularization parameter, respectively. 

\subsection{Implicit Neural Representation Method}
We use INR for the PACT images, i.e., the image is represented as a continuous implicit function by a neural network $\mathcal{M}_{\Theta}$ parameterized with $\Theta$:
\begin{equation}
    \mathcal{M}_{\Theta}:(x, y) \xrightarrow{} \mathcal{H}\label{eq14}
\end{equation}
Image reconstruction with the INR method is then converted to finding the optimal $\Theta$ to minimizing the loss function:
 \begin{equation}
\mathcal{L} =||\mathcal{F}(\mathcal{H})-\textbf{p}_{m}||_{2}^{2}+\eta R(\mathcal{H})\label{eq15}
\end{equation}
Here, $\mathcal{F}$(·) represents the forward operator from heat distribution $\mathcal{H}$ to PA signals (Eq.~\ref{eq4}),  and $\eta$ is a hyper-parameter for the regularization term.
Thus, the network is trained in a self-supervised manner.


\subsubsection{Coordinates Selection}
To train the neural network, PA signals must be related to the spatial coordinate of the PA source.
At a specific sampling moment $t_i$, the PA signal amplitude received by the transducer element at $(x_{0},y_{0})$ is contributed by the point sources locating on the curve $L'(t_i)$ (Fig.\ref{Fig1}b).
After discretization, their coordinate $(x,y)$ can be calculated as:  
\begin{equation}
\left\{
\begin{aligned}
x &= ct_{i}cos(\beta_{0}+j\frac{\alpha}{M-1} + \gamma_{r})+x_{0}  \\
y &= ct_{i}sin(\beta_{0}+j\frac{\alpha}{M-1} + \gamma_{r})+y_{0} 
\end{aligned}
\right.
\end{equation}
where $j \in [0,1, 2, ...M-1]$ indexes the points located on $L'(t_i)$, with $\beta_{0}$ defined as  
\begin{equation}
\beta_{0} = arctan(\frac{y_{0}}{x_{0}})
\end{equation}
and $\gamma_{r}$ is a random angle, following uniform distribution ranging $[-\alpha/(2M-2), \alpha/(2M-2)]$.
\subsubsection{Position Encoding and Network Structure}
After determining the coordinates of the point sources, the coordinates are Hash encoded by multi-resolution mapping\cite{muller2022instant}, and the encoded hash vectors are fed to a MLP, which has two hidden-layers with 128 neurons (Fig.~\ref{Fig1}(c)).
ReLu is selected as the activation function of the neurons and the output activation function is the Sigmoid function to normalize the output value. 
\subsubsection{Training}
During the training process, the initial learning rate was set to 0.001 and was decreased for every 20 epochs with a momentum of 0.5, and Adam optimizer \cite{kingma2014adam} was used to minimize the loss function.
The network was implemented with the tiny-cuda-nn \cite{muller2022instant} framework in Python.

\subsection{Experiments}
\subsubsection{In silico}
A vessel-like structure sized $ 2.5\times2.5$ cm$^{2}$ was placed at the center of a ring transducer array (40 mm radius, 256 elements) in a medium of uniform SoS (1500 $m\cdot s^{-1}$).
The sampling frequency was set at 20 MHz and the size of the computational grids was $512 \times 512$ with a pixel size of 0.05 mm. 
Given the the initial thermal distribution of the vessel-like structure (Fig.~\ref{Fig3}e) and the aforementioned settings, the PA signals were calculated by the forward model (Eq.~\ref{eq12}) presented in Sec. \uppercase\expandafter{\romannumeral2} .

\subsubsection{In vitro}
Five agar phantoms (1$\%$ agar w/v, 6 cm diameter) of different embedding materials forming various shapes were prepared for experimental validation.
Phantom 1  (1$\%$ agar w/v) had three embedded black plastic spheres ($3$ mm diameter) to mimick the scenario of imaging the cross-section of blood vessels in the body.
We also embedded black tungsten wires (0.1 mm diameter) forming leaf branch (Phantom 2), delta (Phantom 3), heart (Phantom 4) and star (Phantom 5) shapes, to mimick the scenario of imaging the longitudinal view of blood vessels in the body.

\begin{figure}
\centerline{\includegraphics[width=\columnwidth]{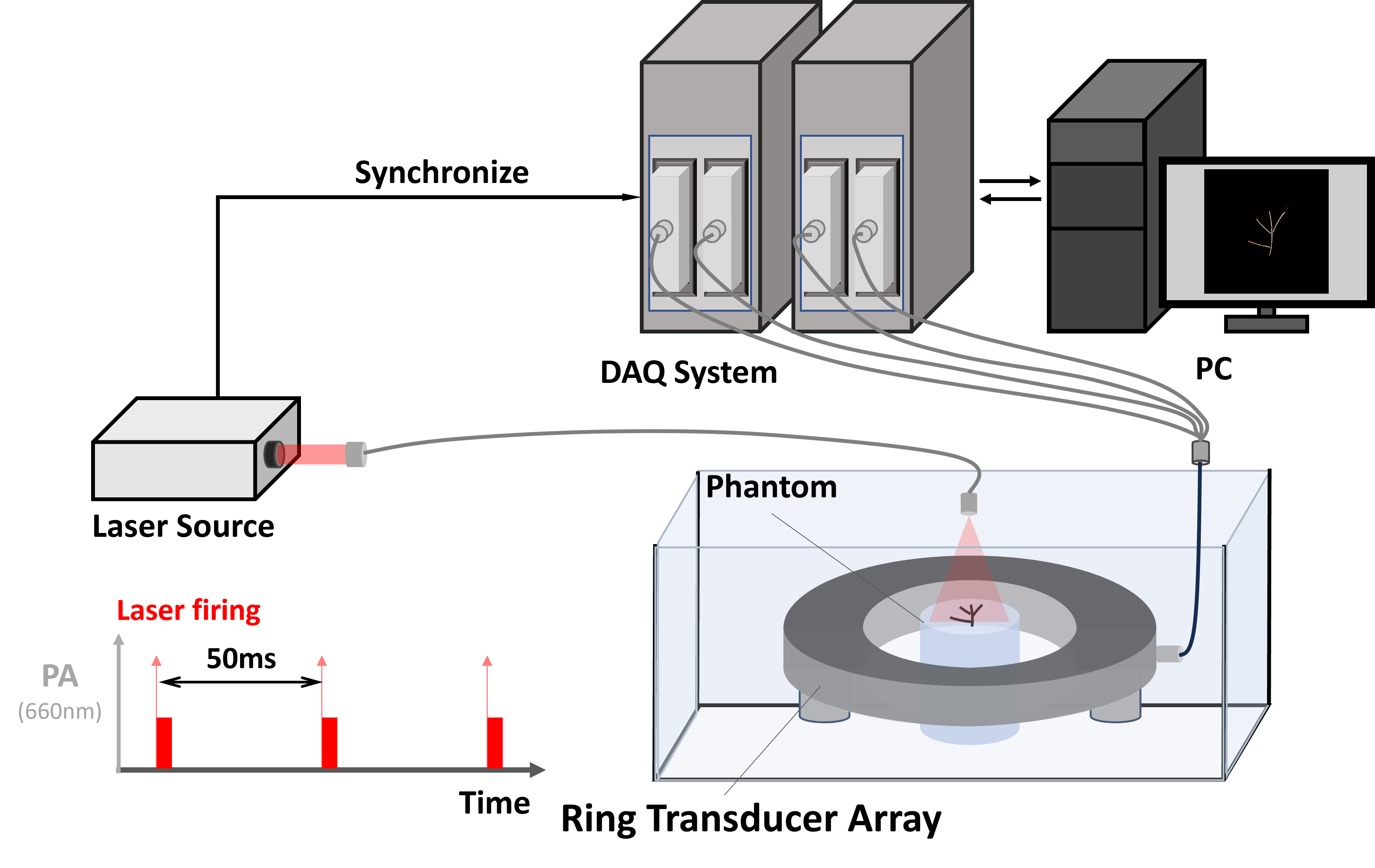}}
\caption{Experimental setup for PACT imaging.}
\label{Fig2}
\end{figure}

A 512-element ring transducer array of 80 mm diameter (center frequency: 5 MHz, Guangzhou Doppler Eletronic Technologies Co., Ltd, Guangzhou, China) was mated with two ultrasound research systems (Vantage 256, Verasonics, Kirkland, USA) for receiving PA signals in the experiments. 
The PA signals (1024 time samples) were initiated by a laser source (PHOCUS MOBILE, OPOTek Inc. USA)  (660 nm wavelength) at 20 Hz repetition rate, synchronizing the data acquisition with 20 MSPS sampling rate.
For all the experiments, the light was emitted from the top-view of the imaging phantom in the de-ionized and degassed water, and the light spot covered the whole ROI (Fig.\ref{Fig2}). The SoS was obtained from the SoS-temperature relationship in pure water\cite{lubbers1998simple}.
During the experiments, the phantoms were place such that the embedded material was in the imaging plane of the ring array.
   
\subsection{Image reconstruction and performance evaluation}
For all the methods, the image reconstruction area was set as $2.5 \times 2.5$ cm$^2$ consisting of $512 \times 512$ pixels.
The signals consisting of 32, 64, 128, 256 and 512 projections were used to reconstruct the images by different methods.  
MB method was implemented following \cite{longo2022disentangling} using non-negative least square optimization method and the iteration number for solving the inverse problem was set to 50.
For the INR method, network training was stopped when the loss function (Eq.~\ref{eq15}) is less than 0.0001.
Total Variation (TV) was used in the loss function of both the MB and INR methods, in which the regularization parameters $\lambda$ and $\eta$ were selected between 0 and 0.2 after trials.
All the above algorithms were implemented in Matlab 2020a on a workstation (Intel Core i7, 14 cores at 2.90 GHz and 16GB memory) interfaced with a graphical processing unit (GPU, Nvidia Geforce RTX 4090 Ti).

For the simulation data, we utilized structural similarity index (SSIM)\cite{wang2004image} and peak signal-to-noise ratio (PSNR) to evaluate the performance of different methods.
For the experimental data, we used signal-to-noise ratio (SNR) and contrast-to-noise ratio (CNR) as the evaluation metrics.

SSIM is defined as:
\begin{equation}
SSIM(f, gt) = \frac{(2\mu_{f}\mu_{gt}+C_{1})(2\sigma_{cov}+C_{2})}{(\mu_{f}^{2}+\mu_{gt}^{2}+C_{1})(\sigma_{f}^{2}+\sigma_{gt}^{2}+C_{2})}
\end{equation}
in which $f$ and $gt$ represent the reconstructed image and ground truth image, respectively. $\mu_{f}$ ($\sigma_{f}$) and $\mu_{gt}$ ($\sigma_{gt}$) are the mean (standard derivation) of image $f$ and image $gt$, respectively, and $\sigma_{cov}$ is the cross-covariance of $f$ and $gt$. The default parameter values of $C_{1}$ is 0.01, and $C_{2}$ is 0.03 \cite{zhou2004Image}, and the dynamic range of these two parameters is from 0 to 1. 

PSNR is defined as:
\begin{equation}
PSNR(f, gt) = 10log_{10}(\frac{I_{max}^{2}}{MSE})
\end{equation}
where $I_{max}$ represents the max value of image $f$ and $gt$, and MSE is calculated by
\begin{equation}
MSE = \frac{1}{n^{2}}||f-gt||_{F}^{2}
\end{equation}
where $||\textbf{·}||_{F}$ is Frobenius norm, and $n$ is the size of image.

SNR (unit in dB) and CNR (unit in dB)  are defined as:
\begin{equation}
SNR = 20 log_{10}(\frac{\overline{I}_{signal}}{\sigma_{background}})
\end{equation}
\begin{equation}
CNR = 20 log_{10}(\frac{|\overline{I}_{signal}-\overline{I}_{background}|}{\sqrt{\sigma_{background}^{2} + \sigma_{signal}^{2}}})
\end{equation}
where $\overline{I}_{signal}$ and $\overline{I}_{background}$ are the average amplitude of the selected imaging object region and background region, respectively, $\sigma_{signal}$ and $\sigma_{background}$ represent the standard deviations (std) of the amplitude in the selected imaging object region and background region, respectively. 
We computed the evaluation metrics with the images reconstructed by different methods for different projection numbers. 

\section{RESULTS}
\begin{figure*}
\centerline{\includegraphics[width=\textwidth]{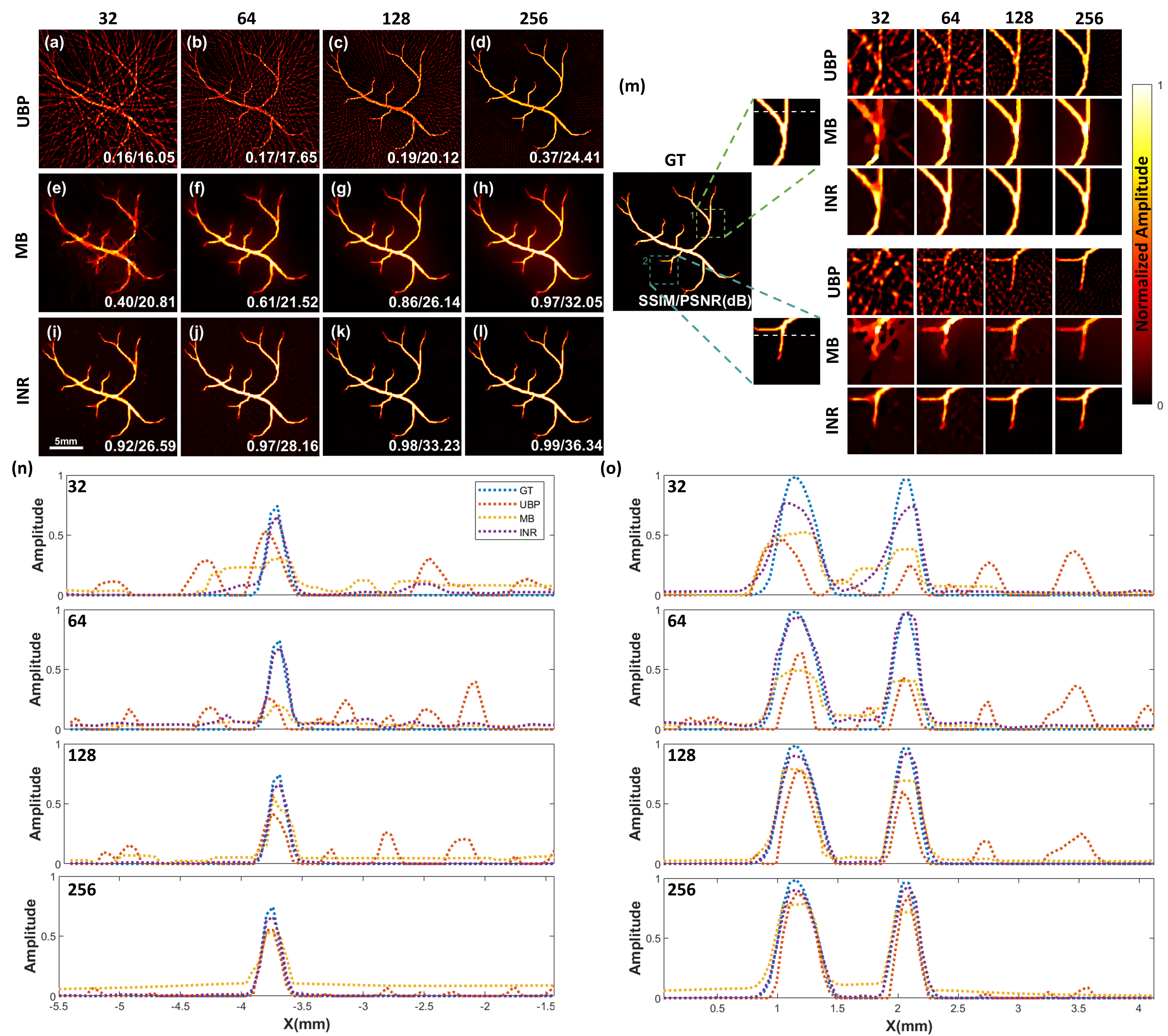}}
\caption{PACT images reconstructed for the simulated data with 32, 64, 128 and 256 projections, (a)-(d) by the UBP method, (e)-(h) the MB method and (i)-(l) the INR method.  (m) Ground truth andzoom in of the two outlined regions (n) and (o) Comparison of the profile lines delineated in outlined regions.}
\label{Fig3}
\end{figure*}

\subsection{Simulation Results}
For the vessel-like structure, the images reconstructed by the UBP, MB and INR methods using 32, 64, 128 and 256 projections are compared in Fig.\ref{Fig3}.
The parameters for the TV regularization term in the loss functions for MB and INR methods were selected as 0.01 and 0.02, respectively.
For the 32, 64, 128 and 256 projections, the network converged after 100, 60, 40, and 20 training epochs, respectively.
The streak-type artifacts are noticeable in the images reconstructed by the UBP method, which became less visible as the projection number increased from 32 to 256 (SSIM/PSNR increased from 0.16/16.05 dB to 0.37/24.41 dB) (Fig.\ref{Fig3}(a-e)).

Compared to UBP, both MB and INR remarkably removed the artifacts and improved the image quality.
As the projection number increased from 32 to 256, the SSIM/PSNR of MB vs INR increased from 0.44/20.81 dB vs 0.92/26.59 dB to 0.97/32.05 dB vs 0.99/36.34 dB (Fig.\ref{Fig3}(f-i, k-o)).
A detailed comparison of the two branches in the vessel-like structure (Fig.\ref{Fig3} e, p, q) and their line-profiles (Fig.\ref{Fig3} (r)-(s)) clearly demonstrated that INR had the best match with the ground-truth data.

\subsection{Phantom Experiments}
For the phantom experiments, the images were reconstructed using 64, 128, 256 and 512 projections by different methods, respectively.
For the MB method, the regularization parameter $\lambda$ was set as 0.05. For the INR method, the signal amplitude was normalized for training the network, and the regularization parameter $\eta$ was set as 0.02. 

\subsubsection{Plastic Spheres}
\begin{figure*}
\centerline{\includegraphics[width=\textwidth]{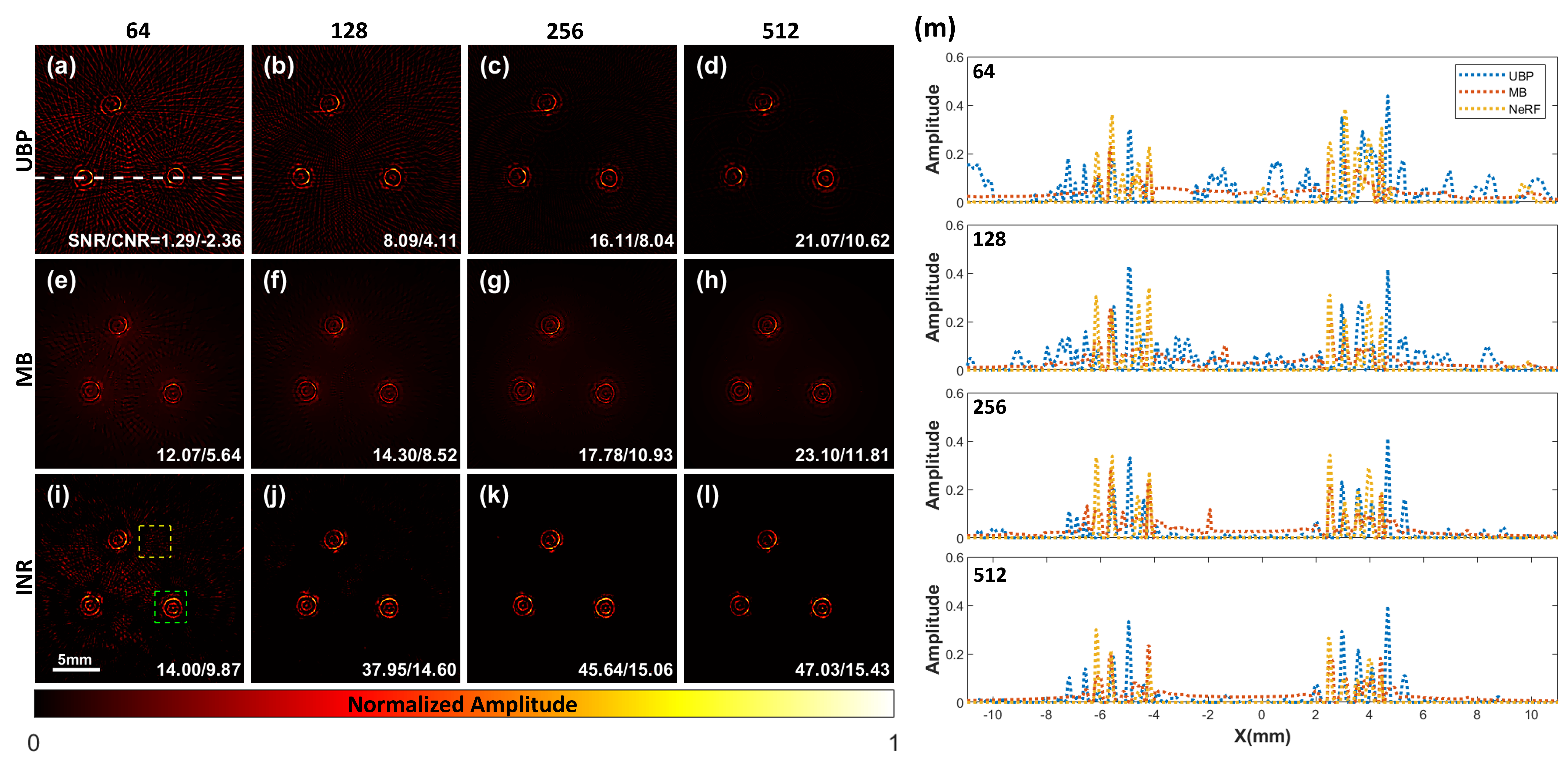}}
\caption{PACT images reconstructed for phantom 1 with 32, 64, 128 and 256 projections, (a)-(d) by the UBP method, (e)-(h) the MB method and (i)-(l) the INR method. (m) Comparison of the profile lines delineated for the same projection number. Green and yellow dash rectangles in (l) represent the signal region and background region for evaluating CNR and SNR.}
\label{Fig4}
\end{figure*}

In phantom 1, multiple rings were observed for each sphere in the images reconstructed by UBP, MB and INR (Fig.~\ref{Fig4}).
Consistent with previous observations in the simulated data, the quality of the images reconstructed by the MB and INR were better than that by the UBP method for the same projection number.
Adding more projection data for image reconstruction also improved the image quality for all different methods (Fig.\ref{Fig4}).
Specifically, the SNR/CNR of the images reconstructed by the UBP, MB and INR were 1.29 dB/-2.36 dB, 12.07 dB/5.64 dB and 14.00 dB/9.87 dB (for 64 projections) and were consistently improved to 21.07 dB/10.62 dB, 23.10 dB/11.81 dB and 47.03 dB/15.43 dB, as the projection number increased to 512 projections.
Noticeably, in the images reconstructed by MB and INR, the two plastic spheres at the lower part in the images each had a dot recovered at the center of the rings, which was not observed in the images by UBP (Fig.\ref{Fig4}(a-l)).
When comparing the profile lines  transversing the two spheres, it was observed that the profile lines in the UBP images had more fluctuations in the background area, while the ones in the MB and INR methods were more flat (Fig.\ref{Fig4}(m)).
Compare to the MB images, the profile lines in the INR images were closer to zero intensity in the background area between the two spheres.

\subsubsection{Tungsten Wires}
\begin{figure*}
\centerline{\includegraphics[width=\textwidth]{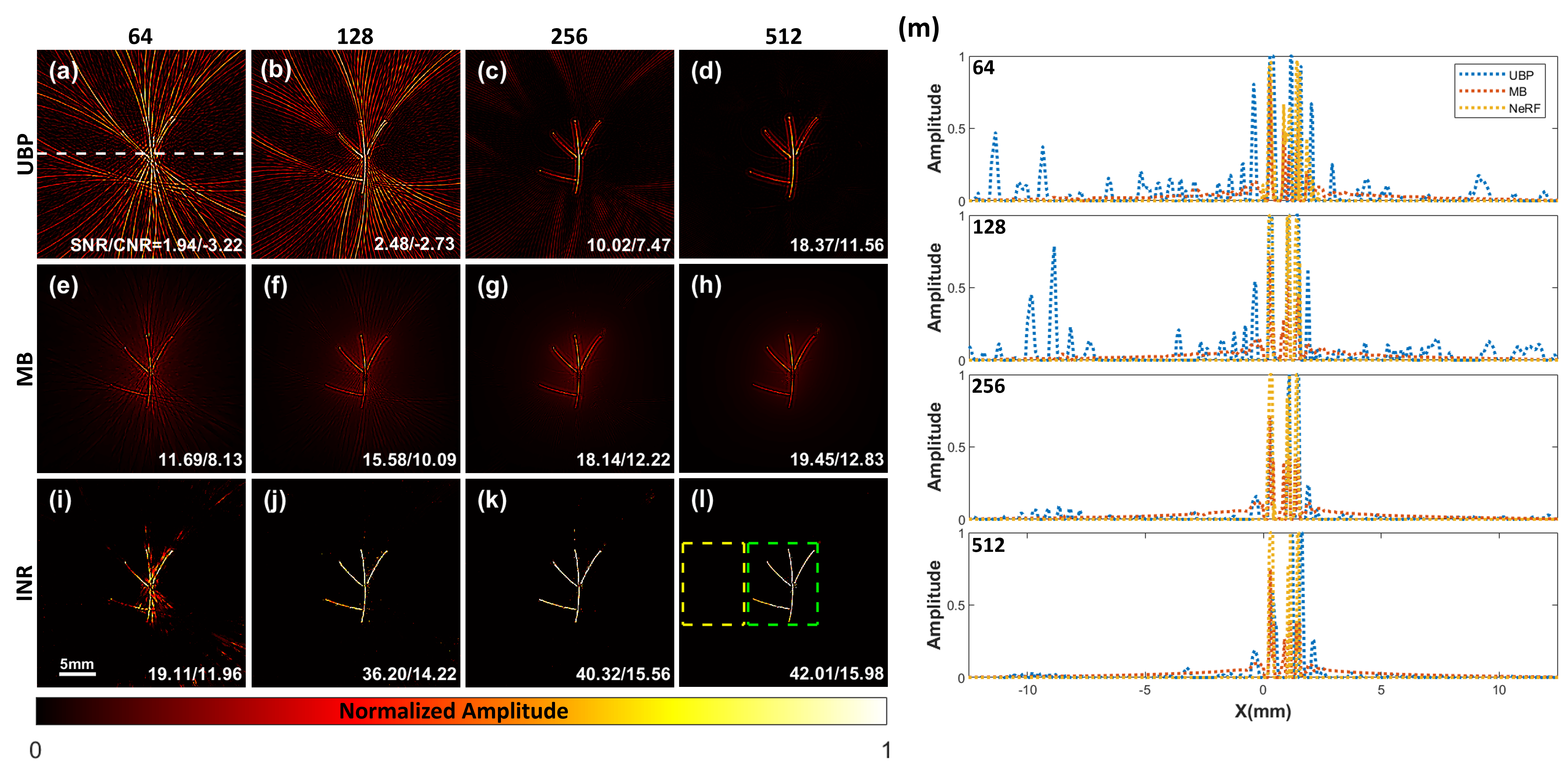}}
\caption{PACT images reconstructed for phantom 2 with 32, 64, 128 and 256 projections, (a-d) by the UBP, (e)-(h) the MB method and (i)-(l) the INR method. (m) Comparison of the profile lines delineated for the same projection number. Green and yellow dash rectangles in (l) represent the signal region and background region for evaluating CNR and SNR.}
\label{Fig5}
\end{figure*}

In phantom 2, the image reconstruction quality of the MB and INR methods was better than that by UBP for the same projection number, which is consistent with the observation in phantom 1. 
For all the methods, the image quality was improved by adding more projection data (Fig.~\ref{Fig5}).
Severe streak-type artifacts in the images reconstructed by UBP when PA signals were very sparsely acquired (64 and 128 projections) (Fig.~\ref{Fig4}(a-b)).
These streak-type artifacts were largely mitigated by the MB and INR methods for the same projection number (Fig.~\ref{Fig4}(e-f, i-j)). 
For 256 and 512 projections, the streak-type artifacts were not obviously observed.
Comparing the profile lines along the middle of the object, it was observed that the ones of MB and INR methods had less fluctuations than that of UBP for all projection number (Fig.~\ref{Fig5}(m)).
Halo artifacts were consistently observed in the middle of the images reconstructed by the MB method, while this artifact was removed in that of the INR method.
Thus, in the background area of images, the profile lines in the INR images were closer to zero intensity than that in the MB images.

Quantitatively, the SNR (CNR) of the images reconstructed by the UBP, MB and INR were 1.94 dB (-3.22 dB), 11.69 dB (8.13 dB) and 19.11 dB (11.96 dB) (for 64 projections) and were consistently improved as the projection number increased, to 18.37 dB (11.56 dB), 19.45 dB (12.83 dB) and 42.01 dB (15.98 dB) (for 512 projections).
For the same number of projections, the SNR (CNR) in the images reconstructed by the INR was consistently better than that of the UBP and MB method, with 17.2--33.8 dB (4.4--16.9 dB) improvements over UBP and 7.5--23.5 dB (3.8--4.1 dB) improvements over MB for 64-512 projections.

\begin{figure}
\centerline{\includegraphics[width=\columnwidth]{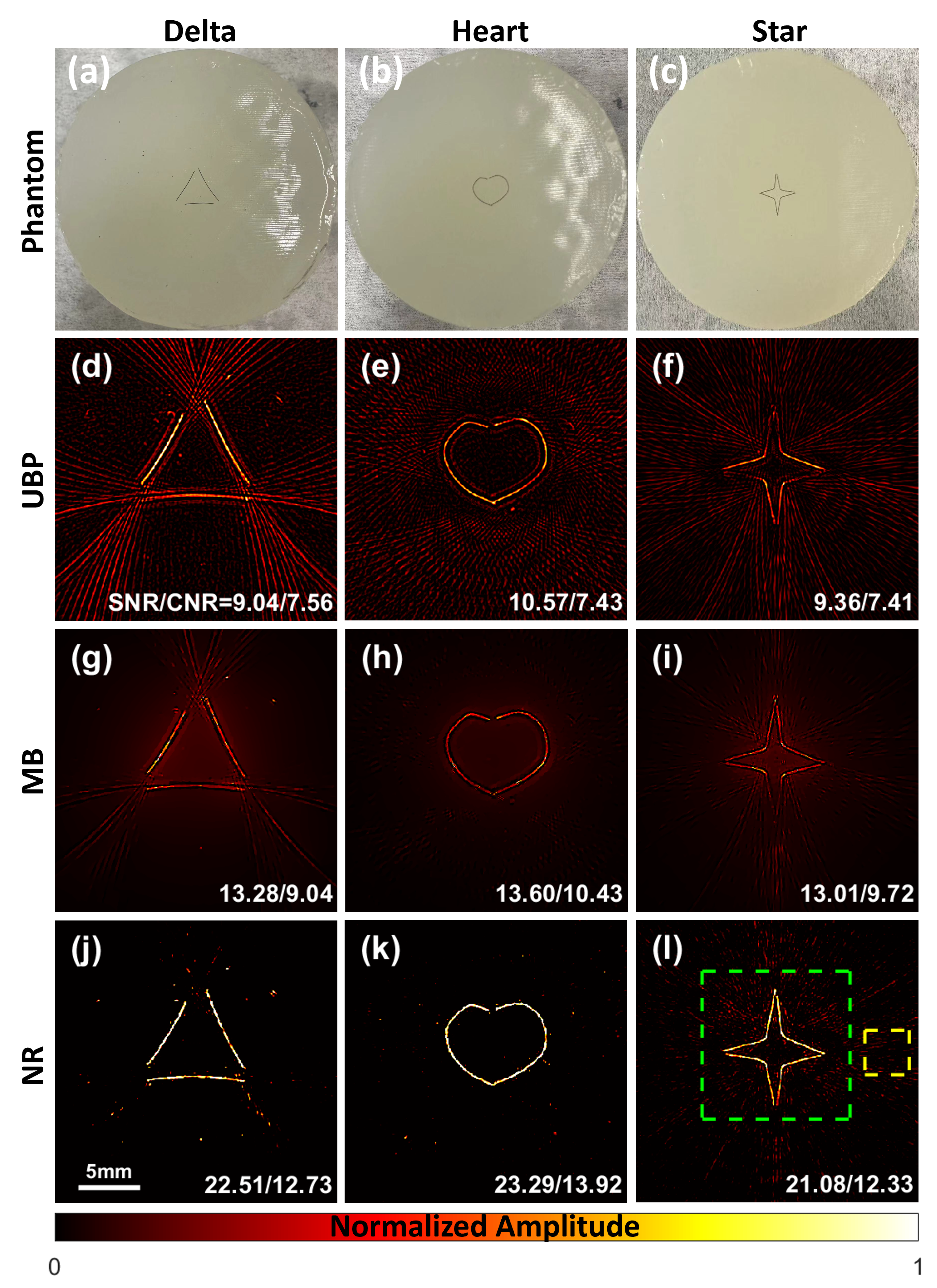}}
\caption{The agar phantoms containing the structure of different shapes: (a) Delta, (b) Heart and (c) Star. (d-) PACT images reconstructed with 128 projections by the (d-f) UBP method, (g-i) MB method and (j-l) INR method.
Green and yellow dash rectangles in (l) represent the signal region and background region for evaluating CNR and SNR.}
\label{Fig6}
\end{figure}

To further demonstrate the superior performance of INR for sparse data, we compared the images reconstructed by UBP, MB and INR for Phantom 3-5 (Delta, Heart, Star) with 128 projections data (Fig.~\ref{Fig6}). For these imaged objects with a radius of $1$ cm, as observed, the images reconstructed by UBP suffered from streak-type artifacts in the background.
With the MB reconstruction, the images had less artifacts for all the phantoms.
With the INR method, most of the artifacts were removed and the fidelity of the object's structure was well maintained in all the phantoms.

Quantitatively, the SNR (CNR) of the images reconstructed by the UBP were 9.04 dB (7.56 dB), 10.57 dB (7.43 dB) and 9.36 dB (7.41 dB) for the phantoms containing the Delta, Heart and Star shaped structures, respectively.
Compared to the UBP method, the SNR(CNR) was improved by 3.03--4.24 dB (1.48--3.00 dB) with the MB method and it was 11.72--13.47 dB (4.92--6.49 dB) with the INR method. 

\section{Discussion}
In this work, we proposed a neural representation framework for image reconstruction in PACT with ring transducer array, to overcome the artifacts and low image fidelity associated with data sparsity.
The proposed method reconstructs the image by learning the continuous representation of the initial heat distribution from raw PA signals directly.
The simulated and in vitro experiments consistently showed that the proposed method outperforms the UBP and MB methods for the same projection number in image quality.
These results demonstrated that the channel count and the hardware cost of a PACT system may be reduced given the proposed neural representation method.

Image reconstruction for PACT is essentially an inverse problem. Analytical methods, such as UBP\cite{b10}, deduce an analytical solution inverting the forward propagation model of ultrasonic wave directly.
Thus, UBP method is conventional for its ease of implementation and low time complexity which provides good image quality when the Nyquist spatial sampling condition is satisfied. This method costed only tens of milliseconds to reconstruct an image in our data, which is in real-time. However, for sparse-view sampling, the image quality of the UBP method is impacted by streak-type artifacts severely.

For the MB and INR methods, the same PA forward model (Eq.\ref{eq4}) was used.
To reconstruct the images, these two methods both solve the inverse problem using optimization techniques to match the model predicted and experimentally measured PA signals.
However, the differences between the two approaches lie in the fact that the former was formulated in a discrete framework while the latter is not.
In the MB method, the initial pressure distribution function is discretized on the image grid and the forward model is represented as a measurement matrix to map the initial pressure to PA signal amplitude received by the transducer element (Eq.~\ref{eq12}), for later calculation of the loss function.
For establishing the measurement matrix, the initial pressure on the arc $L^{'}(t)$ does not necessary falls on the image grid which is often estimated with adjacent pixels on the grid with some interpolation method, which impacts the accuracy and time complexity of the numerical solution \cite{b10, dean2012acceleration}.
In the proposed INR method, the initial pressure distribution function is represented as a 2-D continuous function by a MLP.
Thus, any spatial coordinate (after applying hash encoding) on the traveling path can be fed to the MLP to infer the local initial pressure for relating its contribution to the amplitude of the PA signal.
With the continuous representation, our method interpolates a sampling pixel value from all vertexes of the multi-resolution grids\cite{muller2022instant}, which is more accurate than MB method.
Therefore, compared to the discrete representation of MB method, the continuous representation of INR can better avoid discretization errors, better enforce continuity on the image to mitigate the ill-posed problem and reconstruct high-quality images. 

In terms of the computational cost, the INR method has a better time complexity than MB method as there is no need to calculate the forward model matrix.
For 128 projections, the proposed INR method took 8 minutes to reconstruct the images, while the MB method took 25 minutes in our data (on CPU).
For the time used by the MB method, 9 minutes were spent on the forward model computation for establishing the measurement matrix and 16 minutes were spent on the inversion for image reconstruction.
While the INR method is faster than the MB method, the image reconstruction is not in real-time.
To further expedite the INR method, the photoacoustic forward model can be further simplified, such as expressing the collected RF signals as the summation of point wave source of different wave-forms\cite{SHEN202418}.

Previous learning-based methods, such as AS-Net, PAT-ADN and diffusion model for PACT\cite{guo2022net, song2023sparse,ZHONG2024100613,davoudi2019deep}, applied a pre-trained network model to remove artifacts in sparse-view images.
These methods essentially learned the mapping from low-quality images from sparse-view acquisition to high-quality reference images with densely sampling conditions acquired by a high-channel PACT system.
Thus, these method heavily rely on a mass of high-quality training dataset provided by a high-end system.
Compared to these learning-based methods, our method obtains high-quality PA images from sparse RF signals directly with no need of reconstructing reference images, which reduces the complexity of training model and is free from the limited accessibility to a complex system to build up the training dataset.

Both the in silico and phantom experiments showed the improved image quality with the INR method for the same condition, especially when the used projection number is low.
For all the phantom experiments carried out in this work, as the imaged objects were placed in an area of 1 cm diameter and the wavelength of the array's center frequency was 0.3 mm, the acquisition with a projection number less than 256 was regarded as the sparse-view sampling in the experiments.
In the simulated data, UBP method is easily affected by the streak-type artifacts when the used projection number is low (less than 256).
Thus, the image quality was deteriorated as expected (Fig.~\ref{Fig3}).
With the added prior knowledge provided by TV regularization term, MB method achieved better outcomes than UBP method under the same sampling condition, and improves the SSIM and PSNR of the images remarkably.
Nevertheless, the INR method performed the best under the same sampling condition which has been revealed by both the visual quality of the images, the evaluation metrics (SSIM, PSNR) and the detailed comparison in the sketched profile lines (Fig.~\ref{Fig3}(r)--(s)).
In the experiments, the imaged structures embedded in the phantoms include ball, leaf branch, delta, heart and star shapes.
For phantom 1 (black plastic ball), the sources of PA waves distributed on surface of the plastic balls.
As the imaging plane is estimated of several mm in thickness which is comparable to the diameter (3 mm) of the plastic balls, the PA waves emitted on the surface of different height were all expected being received by the array, resulting in the existence of multiple rings in the reconstructed images.   
For all the phantoms, the INR method can better recover the shape of the structures in the PACT images for the same acquisition condition (Fig.~\ref{Fig4}, Fig.~\ref{Fig5} and Fig.~\ref{Fig6}), with 6.83--19.42 dB (5.85--19.33 dB) improvements in SNR (CNR), compared to the other two methods.
With the improved SNR and CNR, the image fidelity was well recovered and more details were, therefore, revealed in the images reconstructed by the INR method.


In this work, we highlighted the value of the INR approach in solving the image reconstruction inverse problem in PACT with ring transducer array.
The importance of this approach is more prominent under the sparse-view sampling condition.
While we only demonstrated the advantages of the INR method for PACT in homogeneous media, the method can be easily extended to SOS heterogeneous media by additional modification to the forward model, taking the SOS variations of the media into consideration \cite{dean2012acceleration, Huang2013Full-Wave}, which may be important for high quality in-vivo imaging.
In addition, current image reconstruction time of INR method is relatively more expensive than the UBP method, which does not allow real-time imaging speed.
In future work, a more efficient scheme will be explored to reduce the time cost of the INR method.  

\section{Conclusion}
In this work, an INR-based framework for PACT image reconstruction with ring transducer array was proposed, to overcome aliasing-related artifacts associated with sparse-view acquisition conditions.
This framework essentially relies on representing the PA images as a continuous function with a MLP, to better regularize the inverse problem and the network is trained in a self-supervised manner.
In the simulated and phantom experiments, INR method showed superior performance over UBP and MB methods, in preserving image fidelity and in artefacts suppression for the same acquisition condition.
These results clearly demonstrated the value of INR for high-quality PACT image reconstruction with sparse data and its potential in reducing the complexity of PACT systems

\bibliographystyle{IEEEtran}
\vspace{12pt}
\footnotesize
\bibliography{IEEEabrv,Ref}

\end{document}